\documentclass[twocolumn]{aastex62}

\usepackage{longtable}
\usepackage{graphicx}
\usepackage{amsmath,amssymb}
\usepackage{color}
\usepackage{units}
\usepackage{epstopdf}
\usepackage{hyperref}
\usepackage{multirow}

\newcommand{\beq}{\begin{equation}}
\newcommand{\eeq}{\end{equation}}
\newcommand{\bdm}{\begin{displaymath}}
\newcommand{\edm}{\end{displaymath}}

\definecolor{Gray}{gray}{0.9}

\graphicspath{{./plots/}}
\begin{document}

\pacs{95.75.-z,04.30.-w}

\title{The Kitt Peak Electron Multiplying CCD demonstrator}

\author[0000-0002-8262-2924]{Michael W. Coughlin}
\affil{Division of Physics, Math, and Astronomy, California Institute of Technology, Pasadena, CA 91125, USA}
\email{mcoughli@caltech.edu}

\author{Richard G. Dekany}
\affil{Caltech Optical Observatories, California Institute of Technology, Pasadena, CA 91125, USA}

\author{Dmitry A. Duev}
\affil{Division of Physics, Math, and Astronomy, California Institute of Technology, Pasadena, CA 91125, USA}

\author{Michael Feeney}
\affil{Caltech Optical Observatories, California Institute of Technology, Pasadena, CA 91125, USA}

\author[0000-0001-5390-8563]{S. R. Kulkarni}
\affiliation{Division of Physics, Math, and Astronomy, California Institute of Technology, Pasadena, CA 91125, USA}
\affiliation{Caltech Optical Observatories, California Institute of Technology, Pasadena, CA 91125, USA}

\author{Reed Riddle}
\affil{Caltech Optical Observatories, California Institute of Technology, Pasadena, CA 91125, USA}

\author{Tom{\'a}s Ahumada}
\affil{Department of Astronomy, University of Maryland, College Park, MD 20742, USA}

\author{Kevin Burdge}
\affil{Division of Physics, Math, and Astronomy, California Institute of Technology, Pasadena, CA 91125, USA}

\author{Alison M. Dugas}
\affil{Division of Physics, Math, and Astronomy, California Institute of Technology, Pasadena, CA 91125, USA}

\author{Christoffer U. Fremling}
\affil{Division of Physics, Math, and Astronomy, California Institute of Technology, Pasadena, CA 91125, USA}

\author{Gregg Hallinan}
\affil{Division of Physics, Math, and Astronomy, California Institute of Technology, Pasadena, CA 91125, USA}

\author{Thomas A. Prince}
\affil{Division of Physics, Math, and Astronomy, California Institute of Technology, Pasadena, CA 91125, USA}

\author{Jan van Roestel}
\affil{Division of Physics, Math, and Astronomy, California Institute of Technology, Pasadena, CA 91125, USA}

\begin{abstract}
The Kitt Peak Electron Multiplying CCD (EMCCD) demonstrator is a new instrument that has been developed for use at the Kitt Peak National Observatory's 84-inch telescope. The EMCCD enables single-band optical imaging in the Sloan \textit{g} and \textit{r} bands and Johnson \textit{UVRI} filters. The EMCCD is chosen for its sub-electron effective read noise using large multiplicative gains. As a result, frame rates of greater than 1\,Hz are possible. The field-of-view is 4.4$^\prime$ $\times$ 4.4$^\prime$ and the pixel size is 0.259$^{\prime\prime}$. This camera, coupled with a fully roboticized telescope, is ideal for follow-up of short period, white dwarf binary candidates, as well as short duration transient and periodic sources identified by large field-of-view all-sky surveys such as the Zwicky Transient Facility.
\end{abstract}

\section{Introduction}

One of the motivations for the current generation of all-sky surveys are the study of variable and transient sources.
Examples of current and future wide-field telescopes include the Panoramic Survey Telescope and Rapid Response System (Pan-STARRS; \citealt{MoKa2012}), the Asteroid Terrestrial-impact Last Alert System (ATLAS; \citealt{ToDe2018}), the Zwicky Transient Facility (ZTF; \citealt{Graham2018,Bellm2018}), and the Large Synoptic Survey Telescope (LSST; \citealt{Ivezic2014}).
The study of variability, in particular, is an important part of studying stellar and galactic structure and evolution. 
The current generation of astronomical surveys, with a variety of cadences, sensitivities, and fields of view, is resulting in an incredible variety of timescales to study variable objects.
ZTF, in particular, with its $\sim 47$\,square degree field-of-view (FOV) and a range of cadences ($\sim 3$ day -- minutes) is finding variables on timescales from months to minutes. 

The Kitt Peak Electron Multiplying CCD (EMCCD) demonstrator (KPED), consisting of an EMCCD behind a filter wheel imager located at the Cassegrain focus of the Kitt Peak 84-inch (KP84) telescope, is designed for rapid and sensitive photometric follow-up of variable and transient sources identified by these surveys. 
CHIMERA \citep{HaHa2016}, mounted on the Palomar 200-inch and observing simultaneously in two filters, and ULTRACAM \citep{DhMa2007}, mounted on the William Herschel Telescope and observing simultaneously in 3 filters, serve similar purposes.

KPED combines two important technologies: EMCCDs and robotic telescopes. 
EMCCDs use a ``high gain,'' or electron multiplication (EM) register, which is a second stage of readout that amplifies electrons in a process known as ``avalanche multiplication.'' 
The low noise characteristics and rapid readout speeds, allowed by the extremely
low noise ($<$1\,$e^-$ rms) and high-speed clocking of the EMCCDs, makes this mode possible.
At the cost of pixel charge capacity, much higher signal-to-noise ratios (SNRs) are possible above and beyond conventional CCD or CMOS detectors. The large FOV of the instrument is derived from a sensor of 1024 $\times$ 1024 pixels with a plate scale of 0.26 arcsec/pixel, resulting in a 4.4$^\prime$ $\times$ 4.4$^\prime$ FOV. The instrument has sufficient FOV to allow for differential photometry from field stars.

Most EMCCDs use a frame-transfer design, where half of the chip is covered with an opaque layer such that it can serve as a storage unit for the device. The charge is transferred rapidly into the storage era, where it is read out as the next image is taken. This frame-transfer means that observing efficiency is greatly improved. There are some downsides to this technology. For example, the higher frame rates means that subtle noise sources are more visible than in conventional CCDs. There are also EMCCD specific noise sources, such as multiplication noise \citep{TuDh2011}. 
By EMCCD design, a photo-electron entering the EM register can potentially result in a wide range of output signals. For this reason, for a given signal measured, many possible input signals are generally possible. This has the effect of doubling the variance of the signal, which is statistically equivalent to reducing the quantum efficiency of the camera by a factor of two.
Clock-induced-charge, also present in conventional CCDs but often ignored, comes into play for these systems as well.

We now highlight a few science cases for this system.
White dwarf (WD) stars (also known as degenerate dwarfs) are the final stages of evolution of stars with initial masses below approximately 7--9\,$M_\odot$, found in both single and binary systems \citep{DoNa2006}. 
WD binaries, emitting gravitational waves and thereby losing orbital angular momentum, are one of the most common sources for the future Laser Interferometer Space Antenna mission (LISA; \citealt{BrKr2018}).
When these objects merge, they form a variety of exotic objects, including hydrogen-deficient stars such as R Coronae Borealis stars, single subdwarfs, and Type Ia supernovae \citep{IbTu1984,Web1984,SaJe2002}.
Previous measurements of WD binaries have resulted in determinations of the mass-radius relationship \citep{HeBr2014}.
Eclipses can constrain orbital eccentricities, mass ratios of the binaries \citep{Kap2010} and the orbital evolution \citep{FuLa2011}.
Due to their potentially small periods ($\approx 1$\,hr), KPED, with its ability to do rapid and sensitive photometry, is particularly useful for identifying and characterizing WD binaries.
Other periodic sources that can be targeted include aurorae on brown dwarfs, transiting planets, flaring stars, and eclipsing binaries. 

The detection of GW170817 \citep{AbEA2017b} by the Advanced LIGO \citep{aLIGO} and Advanced Virgo \citep{adVirgo} GW detectors ushered in a new era of multi-messenger astronomy when electromagnetic facilities observed a short gamma-ray burst \citep{AbEA2017e} and a kilonova \citep{KiFo2017} in coincidence with the GW detection. To facilitate counterpart detection, many wide-field optical telescopes are expected to image the sky localization regions of GW and Fermi Gamma-ray Burst Monitor (GBM) triggers. Due to the poor localizations, many potential candidates will be detected. For this reason, KPED is well-suited to follow-up of transients identified by other telescopes for the purpose of identification and classification of optical counterparts to GW or short gamma-ray burst (SGRB) events.

The goal for this paper is to present the design and performance of the KPED instrument. Section~\ref{sec:instrument} gives the instrument design. We describe performance of the telescope and camera system in Section~\ref{sec:performance}. We present first light science results from the system in Section~\ref{sec:science}. We present our conclusions and paths for future work in Section~\ref{sec:conclusion}.

\section{Instrument and Software Design}
\label{sec:instrument}

\subsection{Instrument}

\begin{figure}[t]
 \includegraphics[width=3.5in]{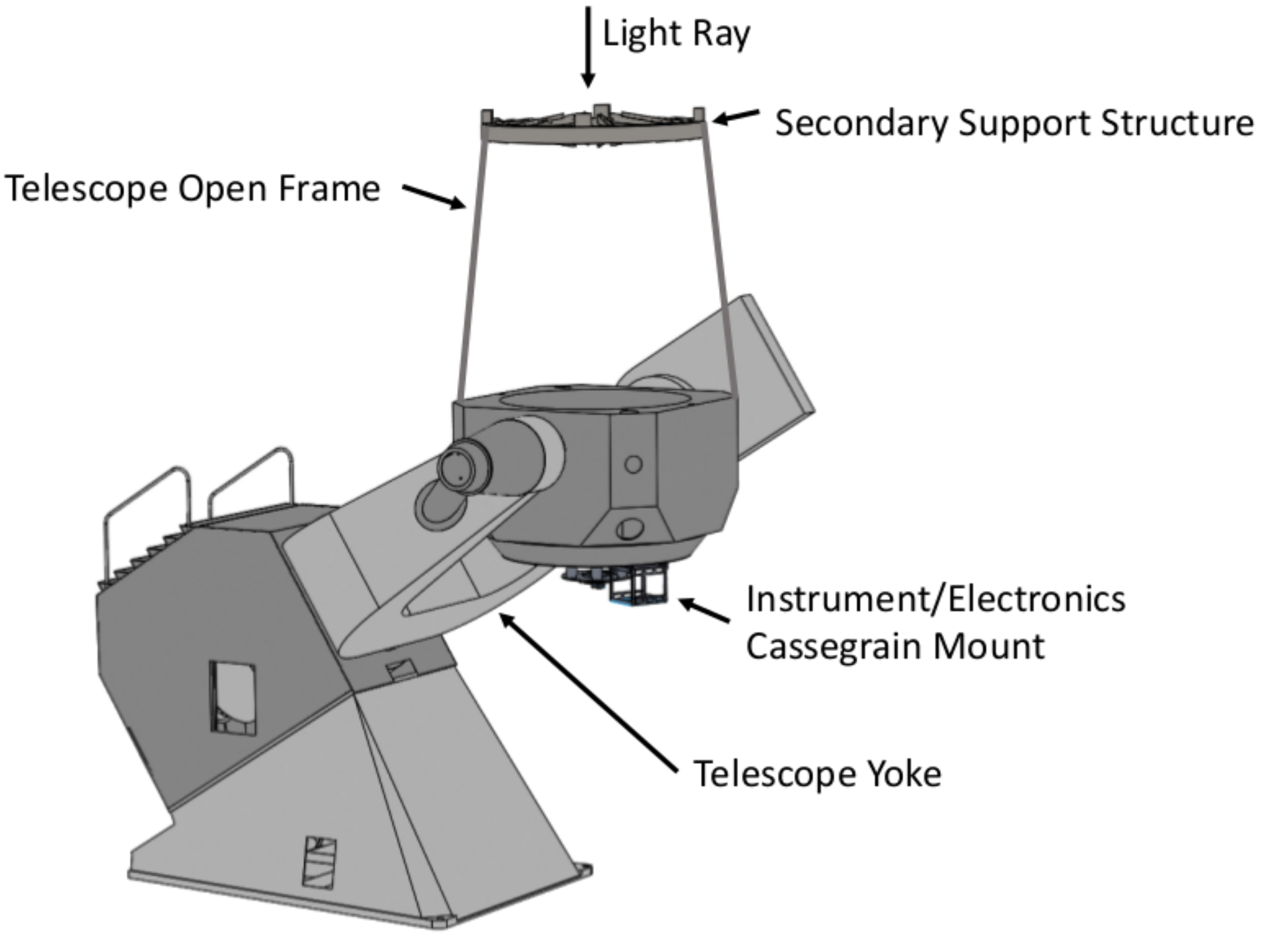}
 \caption{KPED and electronics rack shown in-situ with the 84-inch telescope.}
 \label{fig:2_1m_Telescope}
\end{figure}

KPED is mounted on the backside of the KP84 primary mirror cell with a standoff structure at the Cassegrain $f$/7.6 focal plane (Figure~\ref{fig:2_1m_Telescope}). The standoff structure, adapted from Robo AO \citep{2018AJ....155...32J}, provides a stiff mounting surface in the correct focus position ($\approx$\,0.4\,m from the primary cell). Sitting adjacent to the instrument is a single electronics rack which houses all of the necessary support equipment (e.g.~control computer, network power switch, power/data cables, etc.).
There are two cables to connect the electronics rack to the camera, a USB for communications and control and a Camera Link (Andor) for transferring image data.

\begin{figure*}[t]
 \includegraphics[width=3.5in]{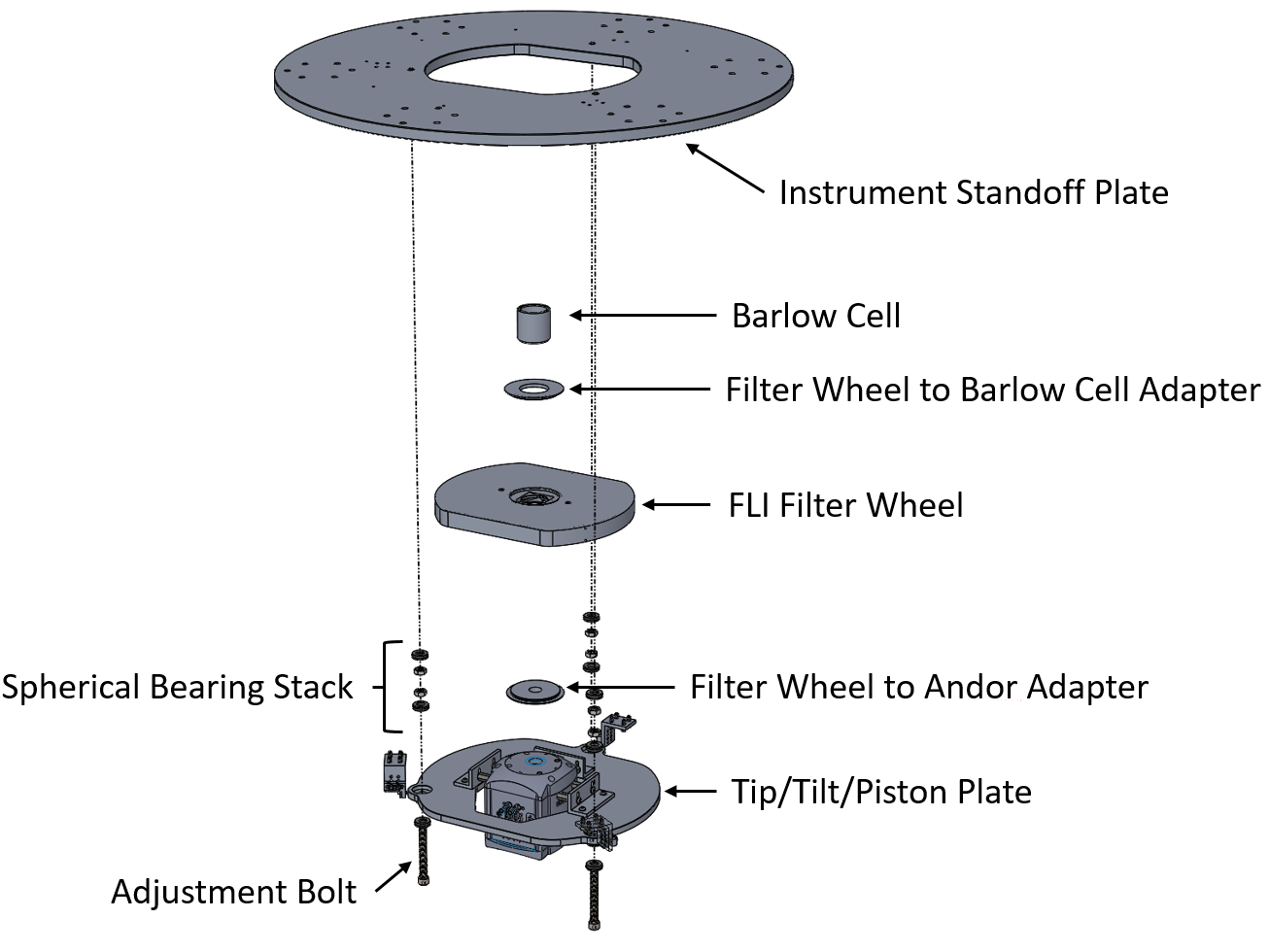}
 \includegraphics[width=3.5in]{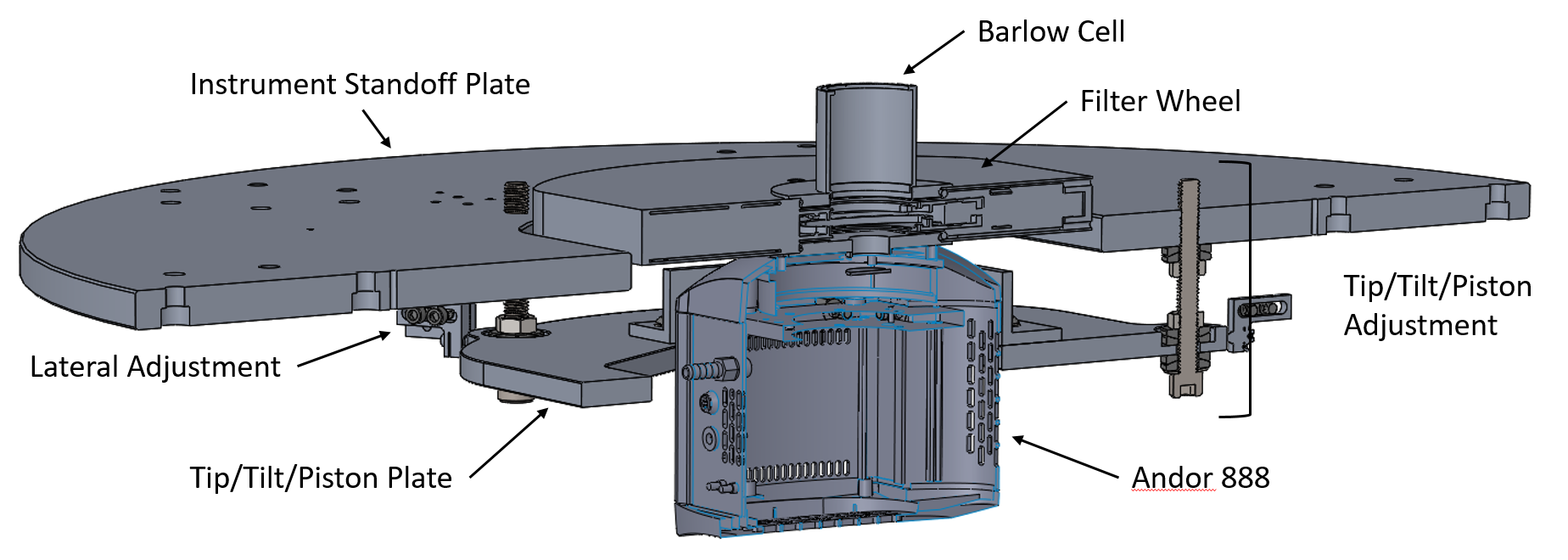} 
 \caption{Instrument diagrams for KPED with the instrument breakdown exploded view on the left panel and the assembled view on the right panel.}
 \label{fig:KPED_Instrument}
\end{figure*}

KPED is comprised of primarily off-the-shelf components and a few custom manufactured parts. The instrument structure is shown on the left panel of Figure~\ref{fig:KPED_Instrument}. A new instrument mounting plate was manufactured for KPED to adapt to the existing instrument standoff structure. This plate provides a rigid mounting surface for the camera assembly and tip/tilt/piston plate. Adjustment of tip, tilt, and piston is achieved by adjusting the length of three 1/2''-13 bolts with a jam nut and spherical washer stack-up. The jam nuts lock the position down and the spherical washers provide a flat, compensating interface. Fine lateral adjustment is achieved through three Newport nudgers at the corresponding bolt interfaces. The right panel of Figure~\ref{fig:KPED_Instrument} shows the instrument in an assembled state.

The camera is an Andor iXon 888 unit; the detector is an array of 1024 $\times$ 1024 pixels with a pixel size of 13\,$\mu$m. The camera is cooled by thermo-electric cooling to reach $-$80$^{\circ}$\,C. With the use of off-the-shelf 1/4''-20 standoffs and custom L-brackets, the iXon 888 is fixed to the tip/tilt/piston plate. The L-brackets are slotted for additional focus adjustment. The front of the iXon 888 has a C-Mount (1''-32 Thread) for mounting various fore optics. KPED utilizes the C-Mount to constrain a Finger Lakes Instruments (FLI) Filter Wheel (part number CFW-3-10) and a Newport PAC087 Barlow lens. 
In the filter wheel, the selectable filters are Sloan \textit{g}, Sloan \textit{r}, and Johnson \textit{UVRI}.
The Sloan filters are the SDSS Gen2 filters from Astrodon. Additionally, there are two empty slots in each of the overlapping wheels for a clear path to the detector and a plastic blank for darks. 
To mate the filter wheel to the camera, a custom C-Mount to the FLI Zero Tilt Mount adapter was manufactured by Precise Parts. On the front face of the filter wheel is a Newport 2''-40 thread lens tube, which houses the Barlow lens. Precise Parts provided a similar adapter to make this connection. 
The light provided from the telescope first enters the PAC087 lens fore optic, then travels through the selected filter, and finally lands onto the iXon 888 detector.

\subsection{Software}

KPED's control software, written in C and compiled with \emph{gcc}, consists of three main tasks.
The first is the telescope control software, the second is the camera control, and the third is the filter wheel control.
All control software is run through terminals, allowing for straightforward remote observations.

The camera control software relies on Andor's software development
kit (2.102.30034).
The software interfaces with the camera through the CameraLink frame grabber PCIe card for image acquisition installed in the control computer.
The camera is run in two modes, one with the conventional amplifier and the other with the EM amplifier.
Within the EMCCD mode, 5 configurations are currently available, with EM gains ranging from 2--200.
Currently, only 1x1 binning is possible, and no sub-framed modes are supported.
The camera buffer is configured such that the oldest image is retrieved first, known as a ``first in, first out'' buffer.

The filter wheel control software relies on Finger Lakes Instrument's software development kit (1.104).
The software requests positions by a numerical index of each of the two filter wheels. 
The numerical index is mapped to the filters using a configuration file.

The telescope control system software handles the remaining requirements for the system. The software governs both the dome and the telescope and is fully integrated into the telescope control software.
Control of telescope is currently given by selection of RA and Declination for sources.
No guiding is currently available.

KPED's data reduction software is based upon the automated data reduction and archiving system of Robo-AO \citep{2018AJ....155...32J}. 
The automated ``house-keeping'' (HK) system uses a distributed task processing queue based on the \texttt{dask.distributed} \texttt{python} module\footnote{\url{https://github.com/dask/distributed}}. The processing results, together with ancillary information on individual observations, are stored in a \texttt{MongoDB} NoSQL database\footnote{\url{https://www.mongodb.com/}}. The whole HK system is containerized using \texttt{Docker} software\footnote{\url{https://www.docker.com/}}.

The data reduction chain for an observing night proceeds as follows. At the end of each night, the EMCCD data are compressed and transferred to the network storage. 
The darks and dome flats taken at the beginning of each night are then combined into master calibration files and applied to the observations.

The next step is the image registration pipeline. The first five frames from an observation are median-combined to avoid selecting cosmic rays as the guide star. The resulting ``master'' reference image is high-pass filtered and windowed about an automatically selected guide star. Each raw short exposure frame is then dark and flat corrected, high-pass filtered, and windowed. We use the \texttt{Image Registration for Astronomy}\footnote{\url{https://github.com/keflavich/image_registration}} package to register the individual frames to the master reference frame, thereby accounting for telescope drift and pointing errors. This is accomplished by using a cross-correlation based sub-pixel image translation registration, followed-by an FFT-based sub-pixel image shift. The pipeline outputs both the FITS-cubes with individually aligned frames as well as a stacked image.

Next, the images are registered to sky coordinates. The astrometric pipeline proceeds as follows. We detect individual sources in the stacked image and extract their pixel positions using the \texttt{SExtractor} software package. We proceed if a sufficient number of sources has been detected. At the moment, we require at least twelve sources. This is more than the 5 independent parameters solved for, corresponding to the RA and Declination of the FOV center, the scale factors in both axes, and the rotation (see Table~\ref{table:astrometry}).
This is currently a conservative choice, and we will explore relaxing this requirement going forward. To identify the sky positions of the extracted sources, we take advantage of the precise astrometry provided by the Gaia Mission Data Release 2 (DR2) catalog \citep{2016A&A...595A...1G, 2018arXiv180409365G}. We query the DR2 catalog within 200$^{\prime\prime}$ of the center of an image (recorded by the 84-inch telescope control system in the ``TELRA'' and ``TELDEC'' FITS header keywords). Next, we cross-match the sources extracted from our image with the Gaia sources. This is accomplished by first generating a coarse a priori astrometric solution to project the Gaia star positions onto the detector plane. We then generate a synthetic image by convolving these projected positions with a synthetic PSF. A synthetic ``detected'' image, generated using the detected pixel coordinates of stars, accordingly centered and padded, is then cross-correlated with the synthetic reference image to provide the offset that is used to cross-match the stars.
The pixel coordinates of a star $i$ in the detector plane (x$_i$, y$_i$) are related to its sky position (RA$_i$, Declination$_i$) via the expressions derived in K. Mierle \& D. W. Hogg (2007)\footnote{\url{http://astrometry.net/svn/trunk/documents/papers/wcs-tutorial/wcs.tex}}. The best fitting solution is found using the bootstrap regression. The solution together with the individual matches are stored in the database. Finally, the resulting FITS World Coordinate System (WCS) is recorded to the registered FITS cubes.

A new data processing pipeline has been developed for photometry. After the standard bias subtraction and flat fielding, photometric light curves of the objects of interest are derived based on differential photometry using field stars. Photometry is performed using both \texttt{SExtractor} \citep{BeAr1996} and \texttt{PythonPhot} \citep{JoSc2015}. In addition, the ability to phase fold lightcurves, important for periodic system, and produce animations from the data cubes is available. 

To enable transient follow-up, image differencing software is available in the standard pipepline.
The pipeline starts by constructing reference images from mosaiced frames from all-sky surveys; there are plug-ins available for SDSS, PS1, and ZTF.
We astrometrically resample these template images using SWarp \citep{BeMe2002} to cover the entire requested field.
SWarp matches the template image to the science image  from  each  epoch  on  a  pixel  by  pixel  level with an alignment typically accurate to $\approx$ 0.1 pixels. 
We use HOTPANTS \citep{Bec2015} to photometrically  match the template image to the science image.
HOTPANTS divides each image into smaller ``stamps,'' and within each stamp, estimates the optimal kernel to convolve the template image into the science image. 
HOTPANTS then outputs a difference image which ideally, outside of transients, should be consistent with zero.

The code is publicly released and available from \texttt{Github} at the links given at the end of the paper.

\section{System Performance}
\label{sec:performance}

\subsection{Image quality}

\begin{figure*}[t]
 \includegraphics[width=3.5in]{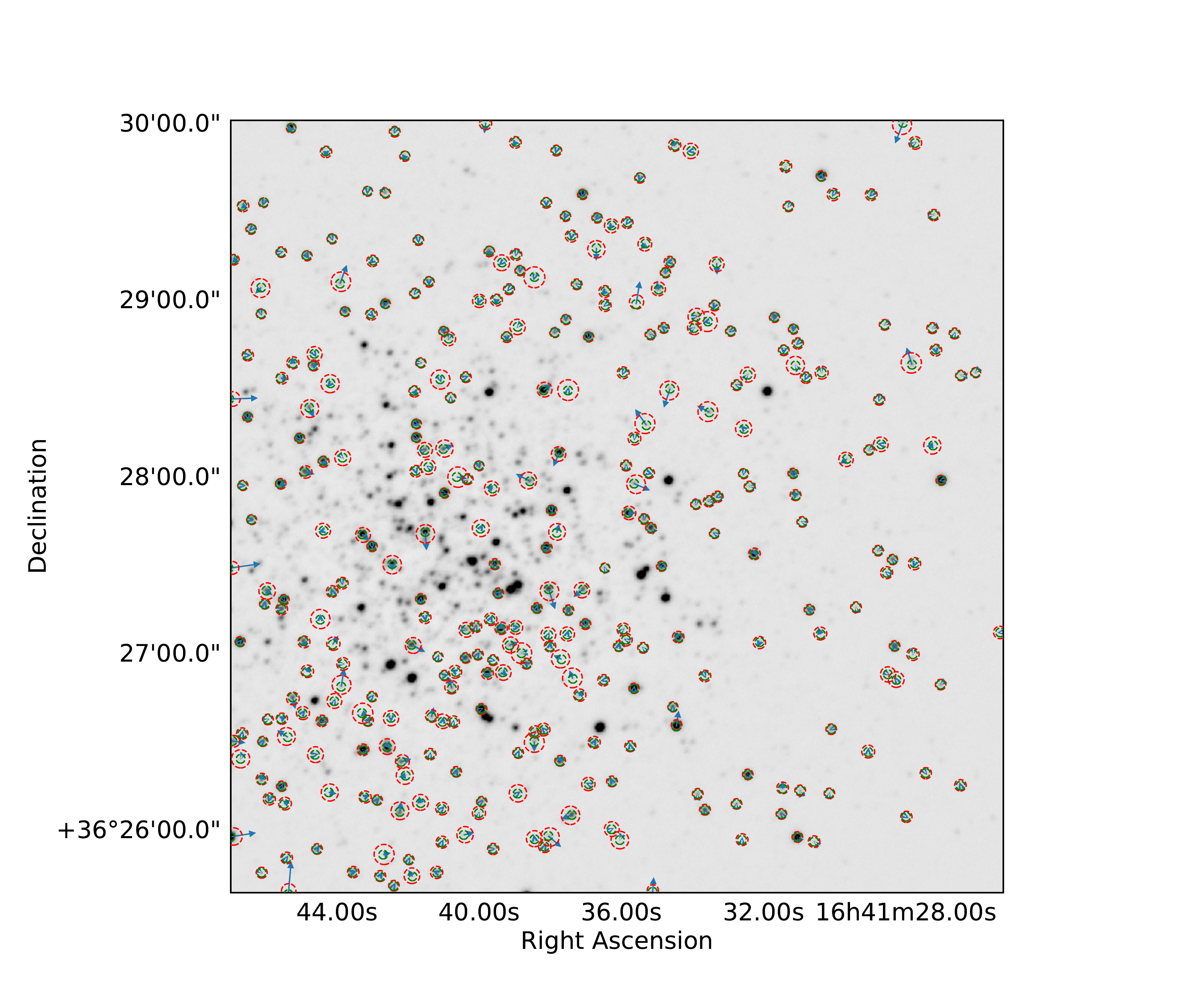}
 \includegraphics[width=3.5in]{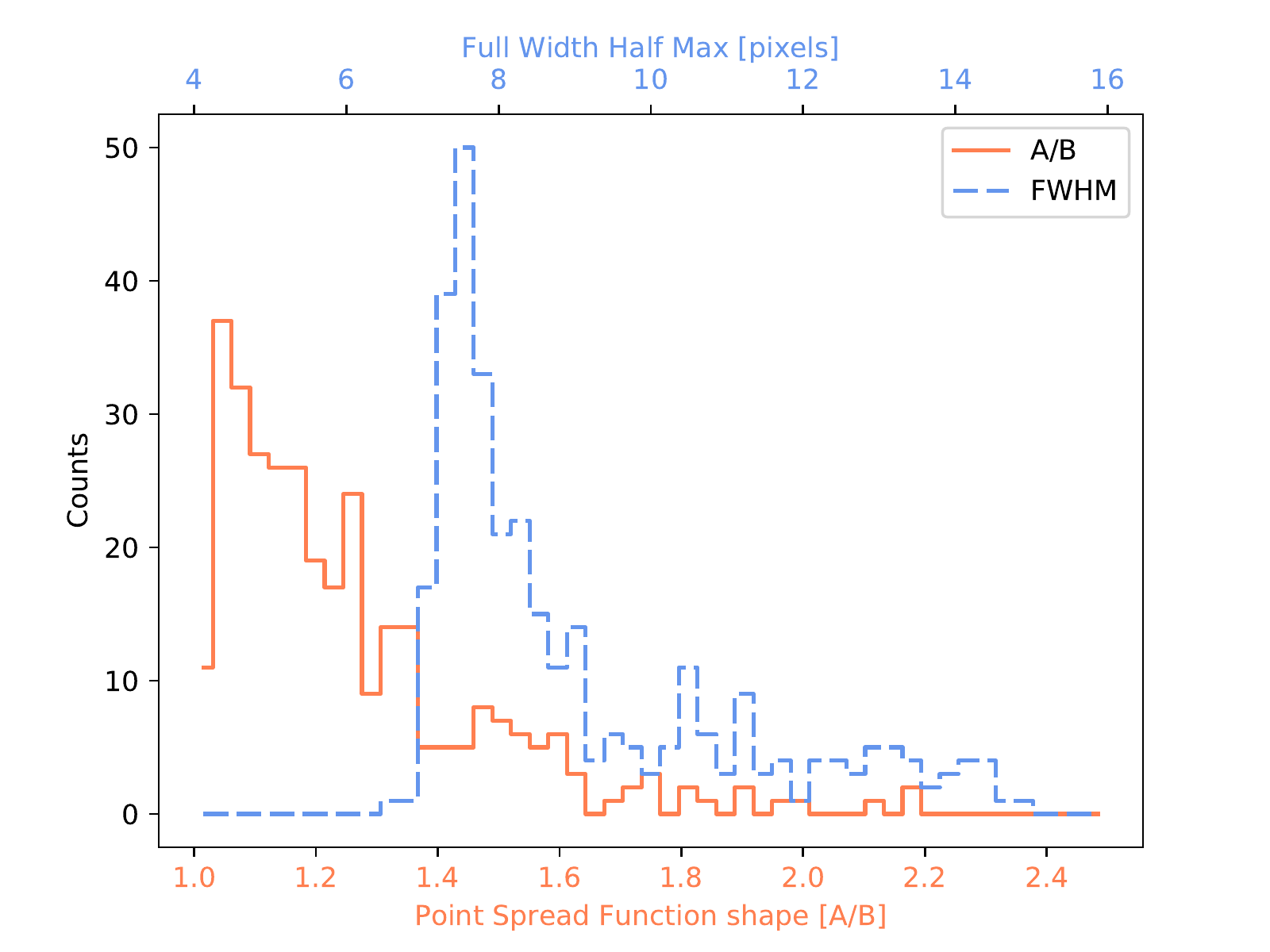}
 \caption{On the left is a KPED image of M13 in \textit{r}-band. The green and red circles correspond to PS1 and KPED matched sources identified in the image. Blue arrows, 10$\times$ the size of the true separation, are also shown. On the right is a pair of histograms, both the point spread function shapes (A/B) and the full-width half maximum of the point spread function.}
 \label{fig:M13}
\end{figure*}

\begin{figure}[t]
 \includegraphics[width=3.5in]{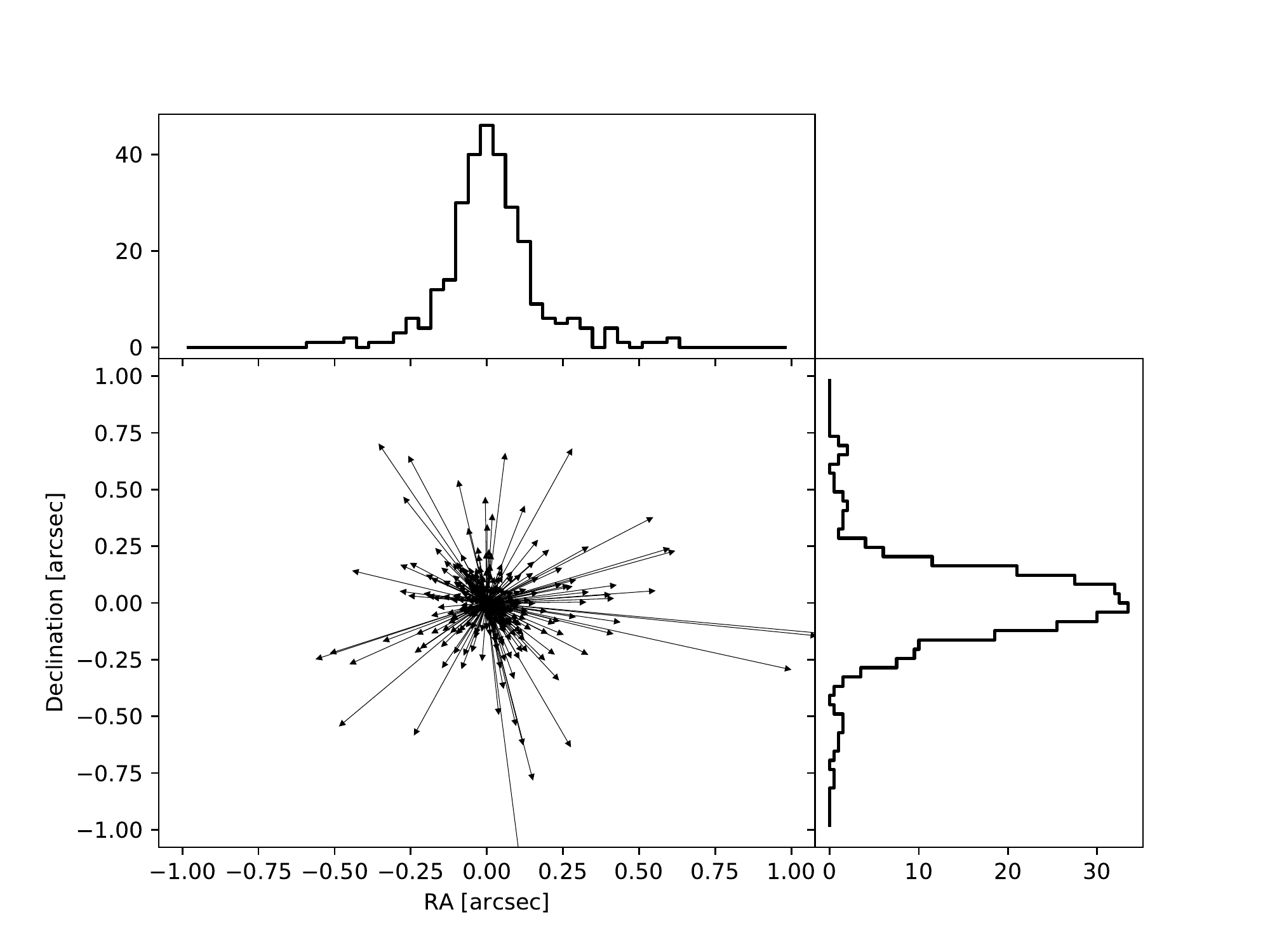}
 \caption{Quiver plot showing the distortion measured in the M13 image.}
 \label{fig:M13_quiver}
\end{figure}

To assess the image quality, we took images of Messier 13 (see left panel of Figure~\ref{fig:M13}), also known as NGC 6205 or the Hercules globular cluster (RA=16:41:41.24, Declination=+36:27:35.5). On the right of Figure~\ref{fig:M13} is a pair of histograms, both the point spread function shapes (A/B) and the full-width half maximum (which assumes a Gaussian core) of the point spread function. 
The point spread function shape is represented by an ellipse with the lengths of the semi-major and semi-minor axes denoted as A and B respectively. The histograms indicate that the shape of the point spread functions are approximately circular, with the shapes A/B peaking near to 1. In addition, the peak of the full-width half maximum of the point spread function peaks between 6--8\,pixels. At our plate scale, this indicates a seeing of approximately 1.5--2$^{\prime\prime}$, consistent with that expected at the 84-inch \citep{2018AJ....155...32J}. To determine the amount of distortion in the field, we use the PS1 catalog to match with KPED sources. In Figure~\ref{fig:M13_quiver}, a quiver plot shows the distortion measured in the M13 image. Separations are typically less than 0.5$^{\prime\prime}$, indicating excellent astrometry across the field.

\subsection{Pointing}

We generated a pointing model using 120 observations of bright Gaia DR1 stars ($6 \leq \mathrm{G_{mag}} \leq7$).
This was performed by first griding the sky in azimuth and elevation such that 120 points spread uniformly on the sky above an elevation of 20$^\circ$ were chosen. 
To minimize the time required to image the sources, a greedy path was constructed to minimize the slew distances. 
During these observations, the telescope was first slewed to the coordinates of a star. At that point, if the star was identified in the field, the operator offset to the location of the star, and the successful find and its offset was noted. If not, the lack of success was noted, and the process was repeated for the next star.
After the pointing model was produced from this data; the ``blind'' pointing errors are within 1\,arcminute, more than sufficient given the FOV of the instrument.

\subsection{Photometric Performance}
\label{subsec:photometry}

\begin{figure}[t]
 \includegraphics[width=3.5in]{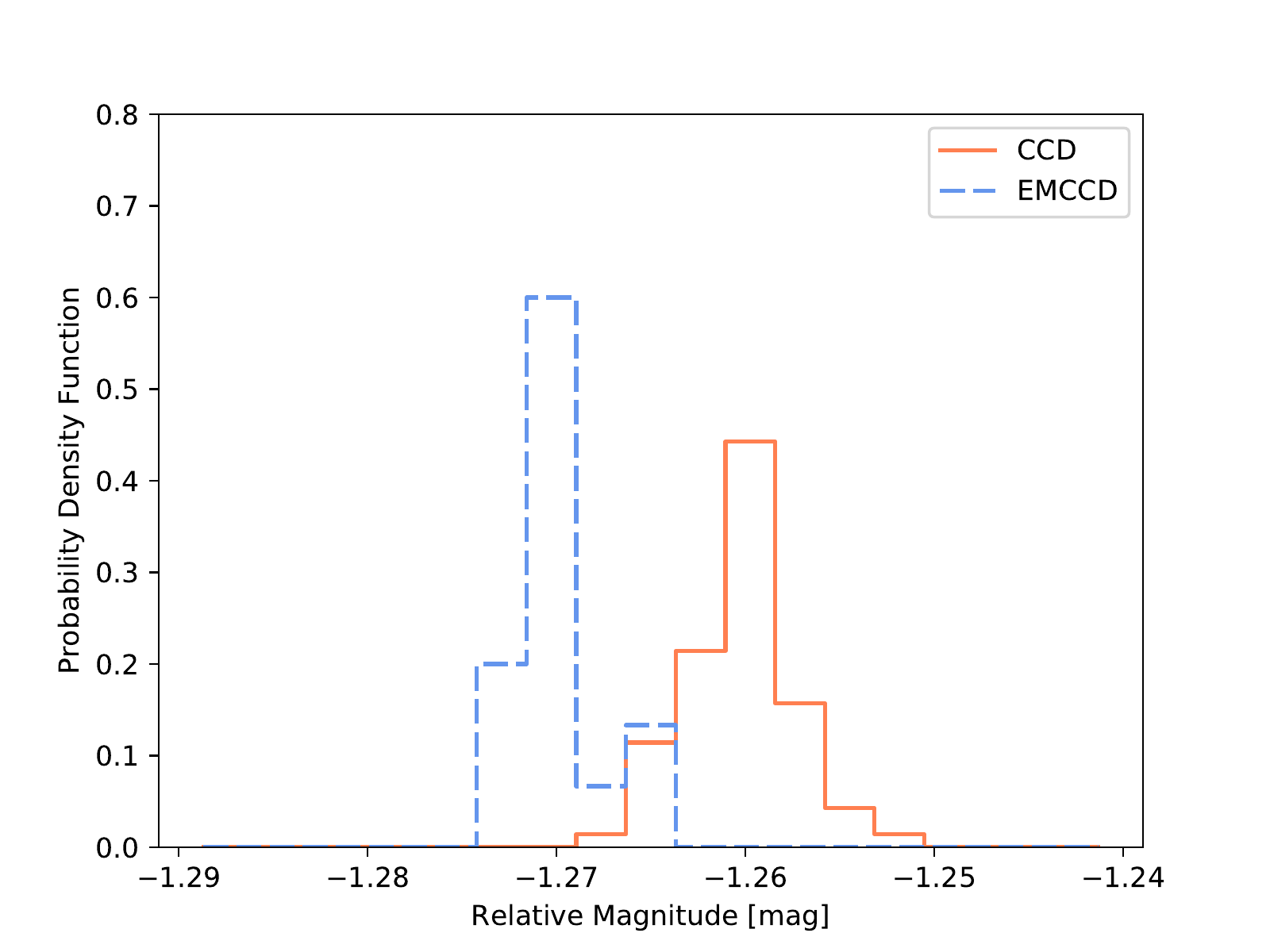}
 \caption{Photometry for Hz44 
 Photometry of HZ44 using both the conventional and EMCCD modes. The modes are consistent at the 0.01\,mag level.}
 \label{fig:photometry}
\end{figure}

To assess both the photometric stability and compare between the conventional CCD and EMCCD modes in the camera, we took a series of images of Hz44, an European Southern Observatory (ESO) standard star (RA=13:23:35.37, Declination=+36:08:00.0). It has a \textit{V}-band magnitude of about 11.7, which is bright enough for the statistical errors on the photometry to be small ($\textrm{mag}_\textrm{error}<0.01$). By this comparison, the photometry of HZ44 using both the conventional and EMCCD modes is consistent to within 0.01\,mag (Figure~\ref{fig:photometry}).

\subsection{Tracking}

\begin{figure}[t]
 \includegraphics[width=3.5in]{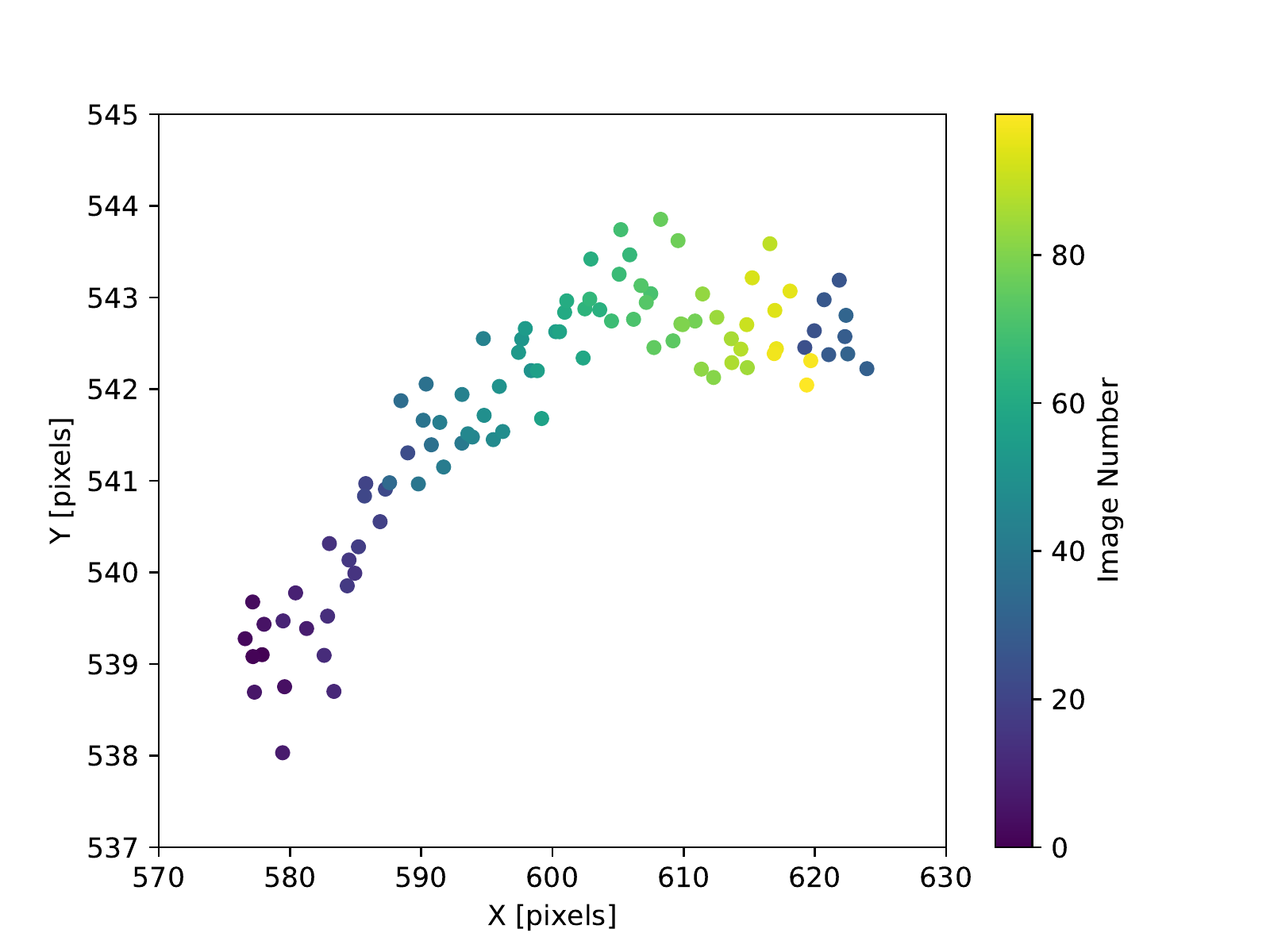}
 \caption{X and Y pixel positions as a function of time for Hz44 in the EMCCD mode. The drift in X (corresponding to right ascension) is significantly larger than the drift in Y (Declination).}
 \label{fig:guiding}
\end{figure}

To assess the performance of the tracking system, we use the same dataset from Section~\ref{subsec:photometry} and explore the position of the star as a function of time over the hour of data. Figure~\ref{fig:guiding} shows the X and Y pixel positions as a function of image number over the hour of data. The drift is evident, and predominantly in right ascension (aligned with the X-axis). Over an hour, the star drifts about 40 pixels, which is approximately 9.6 arcseconds per hour. This will be resolved in the future by adjusting the RA drive frequency.

\subsection{Astrometric solution}

The mean astrometric solution for KPED obtained using the best 36 successfully solved fields is given in Table \ref{table:astrometry}. This solution is used as the default in the header files of KPED data before further reductions. Reduced data uses this solution as the basis for further refinement based on identified objects in the field.

\begin{deluxetable}{cccc}
\tablehead{\colhead{FITS WCS keyword} & \colhead{Value} & \colhead{Standard Deviation} & \colhead{Unit}} 
\startdata
CD1\_1           & $-$7.120$\times 10^{-5}$ & 6$\times 10^{-8}$   & deg/pix    \\
CD1\_2           & 5.0$\times 10^{-7}$    & 8$\times 10^{-8}$  & deg/pix    \\
CD2\_1           & 5.3$\times 10^{-7}$    & 6$\times 10^{-8}$   & deg/pix    \\
CD2\_2           & 7.121$\times 10^{-5}$  & 8$\times 10^{-8}$   & deg/pix    \\
CRPIX1           & 512.019   & 0.003  & pix        \\
CRPIX2           & 512.002   & 0.002  & pix        \\
CCD\_ROT         & 0.423     & 0.045  & deg        \\
PIXEL\_SCALE1    & 0.2563    & 0.0002 & arcsec/pix \\
PIXEL\_SCALE2    & 0.2563    & 0.0003 & arcsec/pix \\ \hline
\enddata
\caption{KPED astrometric solution.}
\label{table:astrometry}
\end{deluxetable}

\subsection{Zero point and sensitivity}
For transient observations where absolute magnitudes are important, the zero point of the field is calibrated using PAN-STARRS1 and Sloan Digital Sky Survey (SDSS) objects as standards. Given the coordinates of the target, an on-the-fly query to PAN-STARRS1 and SDSS retrieves the stars within the field that have a minimum of 4 detections in each band. To obtain the zero point of images taken with the \textit{g} or \textit{r} filters, a direct comparison is made between the stars in the field and the standards in the retrieved catalogs. However, for the  Johnson \textit{U} and \textit{I} bands, which are different from the PAN-STARRS1/SDSS \textit{u} and \textit{i} bands, the transformations of \citealt{Jo06} are applied. Table \ref{table:sensitivity} summarizes sensitivities for the commonly used bands for KPED with 300\,s exposures, which is the standard adopted for transient observations.

\begin{deluxetable}{cccc}
\tablehead{\colhead{Band} & \colhead{Limiting mag.} & \colhead{$\sigma_{\rm Limiting \hspace{0.25em} mag.}$} & \colhead{Zeropoint mag.}}
\startdata
\textit{r} & 23.3 & 0.31 & 20.8\\
\textit{g} & 23.7 & 0.48 & 21.3\\
\textit{U} & 17.5 & 0.99 & 17.0\\
\textit{I} & 16.2 & 0.49 & 16.1\\
 \hline
\enddata
\caption{KPED sensitivity and zero points for the bands used for transient observations. The 5\,$\sigma$ limiting magnitudes are quoted for 300\,s observations (except for \textit{I}-band, which is 5\,s due to image saturations), which is the standard exposure time used.}
\label{table:sensitivity}
\end{deluxetable}

\section{Science Results}
\label{sec:science}

In this section, we present examples of three science cases described above: the characterization of eclipsing systems, pulsators, and photometry of ZTF transients.

\subsection{Eclipsing systems}

\begin{figure}[t]
 \includegraphics[width=3.5in]{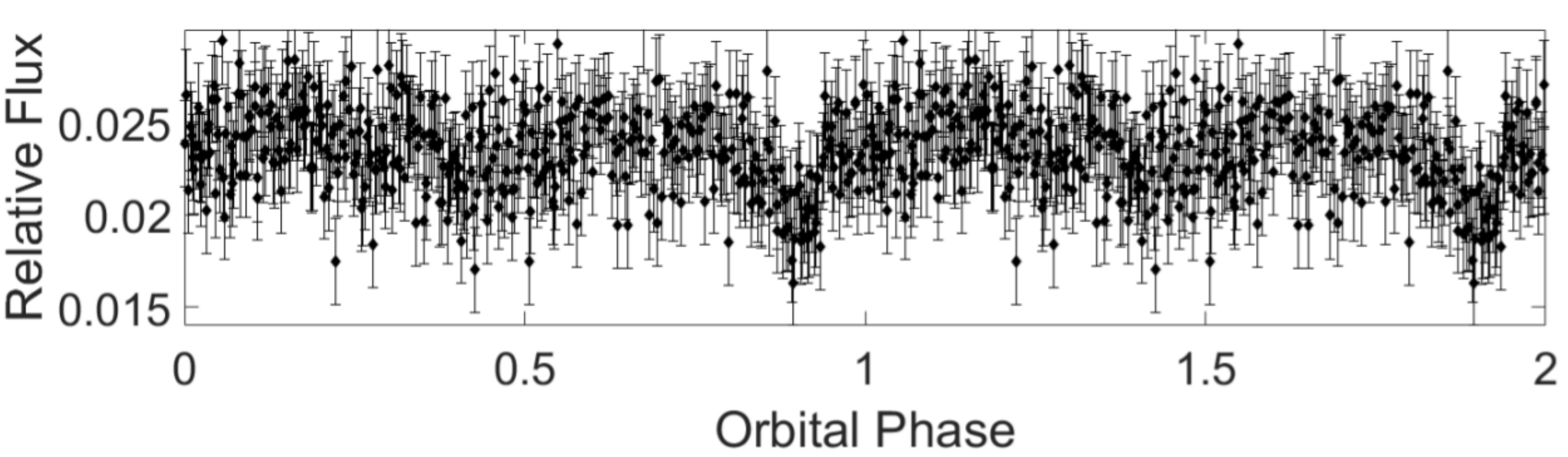}
 \caption{\textit{g}-band lightcurve of J0651, folded to a period of 12.75\,minutes. Each data point corresponds to a 1-s integration. The data are reflected so as to showcase two identical cycles.}
 \label{fig:J0651}
\end{figure}

To highlight the rapid cadence of lightcurves possible with KPED, we begin with observations of SDSS J065133.338+284423.37, known as J0651, which is a detached WD binary with a 12.75\,minute period \citep{BrKi2011,HeKi2012}.
It is the shortest-period detached compact binary currently known.
J0651 is special also for having both primary and secondary eclipses, ellipsoidal variations and Doppler
boosting, providing photometric markers whereby all orbital and system parameters can be measured.
As a short-period system, it is useful for tracking the orbital decay due to gravitational wave radiation \citep{HeKi2012}, and one of the strongest known sources for LISA \citep{KuKo2018}.
Figure \ref{fig:J0651} shows a 1-s cadence lightcurve of J0651, folded to a period of 12.75\,minutes, highlighting the primary and secondary eclipse for the system.

\begin{figure}[t]
 \includegraphics[width=3.5in]{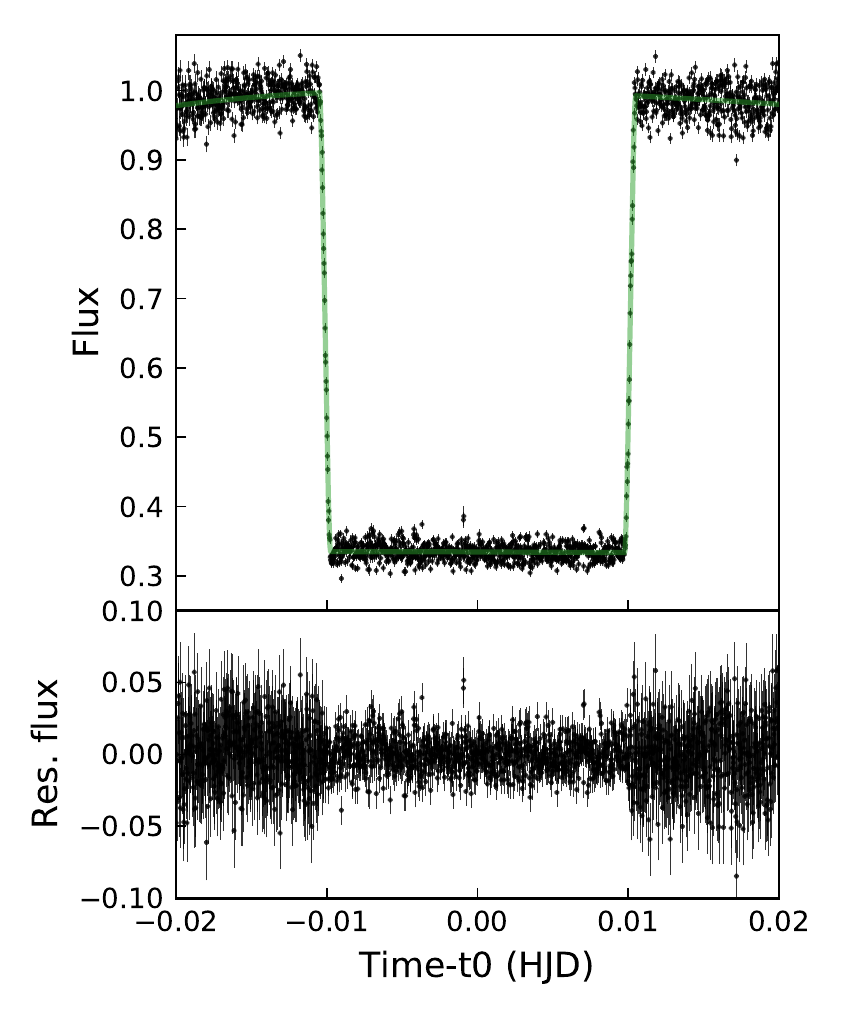}
 \caption{\textit{r}-band lightcurve of PTF1J162528.61$-$003545.8, showing the primary eclipse with the best-fit model overplotted. Each data point corresponds to a 1-s integration.}
 \label{fig:PG1336}
\end{figure}

To show the potential use of high-cadence lightcurves for modelling, we observed the primary eclipse of PTF1J162528.61$-$003545.8, an eclipsing white dwarf--red dwarf system ($g=16.0$) identified by PTF with an orbital period of 7.8\,hr (Figure \ref{fig:PG1336}). We obtained a one hour light curve which covers the primary eclipse. Data were taken at 8\,Hz and stacked to 1\,s images. We modelled the lightcurve using \textsc{ellc} \citep{2016A&A...591A.111M} and \textsc{emcee} \citep{2013PASP..125..306F}. This showed that we can time the eclipse with a statistical uncertainty of 0.20\,s and the sum of the scaled radii ($[R_1+R_2]/a$) and the ratio of the radii with a precision of $\approx 1\%$. The typical uncertainty on the individual measurements was 2.5\%.

\subsection{Ultrashort Period Pulsating stars}

\begin{figure}[t]
 \includegraphics[width=3.5in]{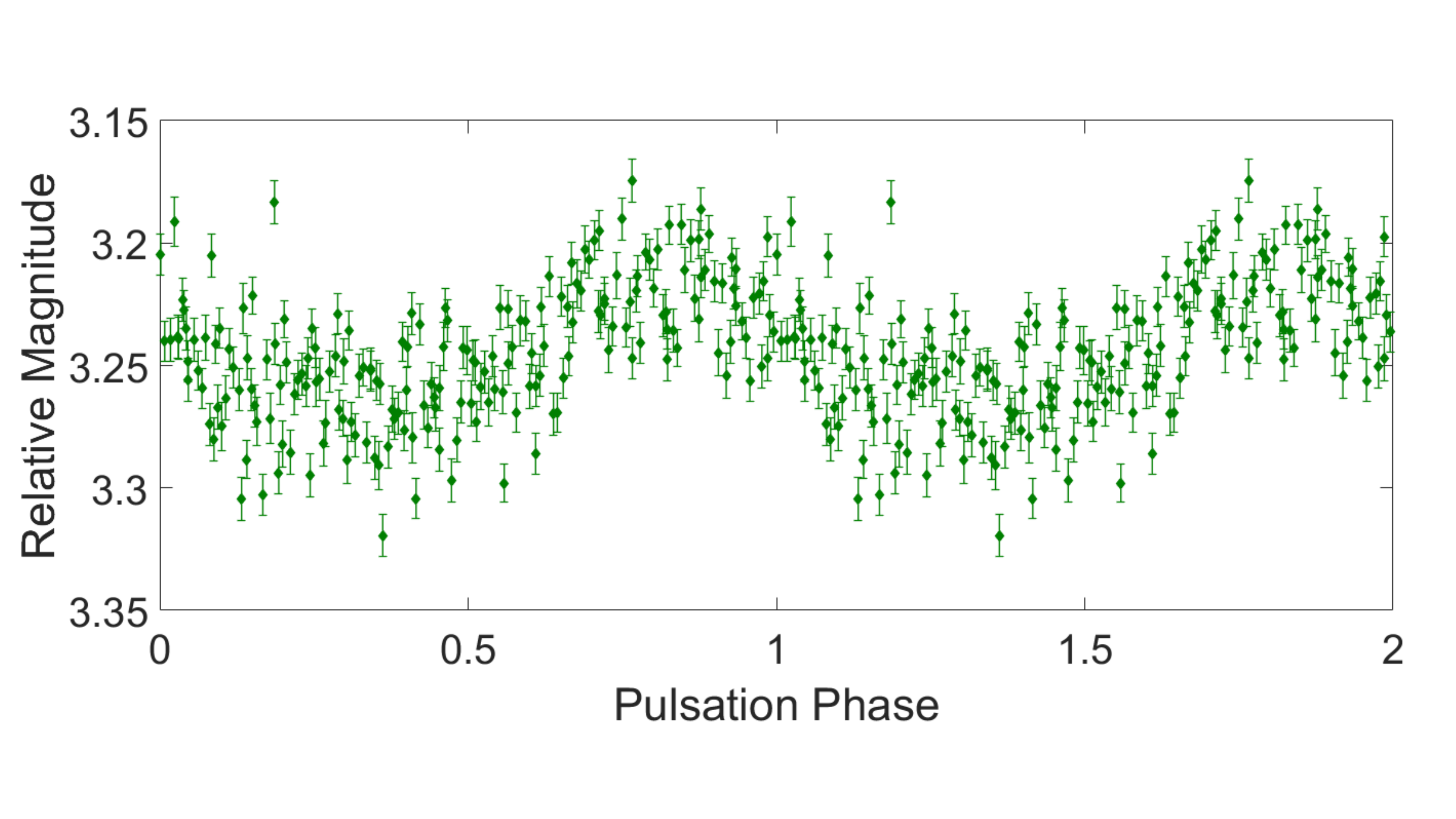}
 \caption{\textit{g}-band lightcurve of PTF1J214022.55+262124.4 folded to a period of 4.8\,minutes. Each data point corresponds to a 1\,s integration. The data are reflected so as to showcase two identical cycles.}
 \label{fig:1821I}
\end{figure}
One key advantage of an EMCCD detector lies in its ability to effectively eliminate readout times--enabling short exposures without sacrificing a substantial fraction of time to read out. Thus, an ideal science case for such an instrument is photometrically following up objects which require a rapid cadence of exposures because they exhibit variable behavior on very short timescales. One such class of objects are pulsating white dwarf and hot subdwarf stars, which typically exhibit periods on the order of 5 minutes. To demonstrate KPED's capability in recovering the period waveform of one such object, we observed a newly discovered DB White Dwarf pulsator PTF1J214022.55+262124.4. Figure \ref{fig:1821I} illustrates the phase folded lightcurve of the object, which clearly demonstrates the detection of the fundamental pulsation mode, which exhibits a period of only 4.8 minutes. Pulsators serve as particularly interesting objects of study because pulsations can allow us to probe the interior structure of objects such as white dwarfs via astroseismology. 

\subsection{ZTF Transients}

\begin{figure}[t]
 \includegraphics[width=3.5in]{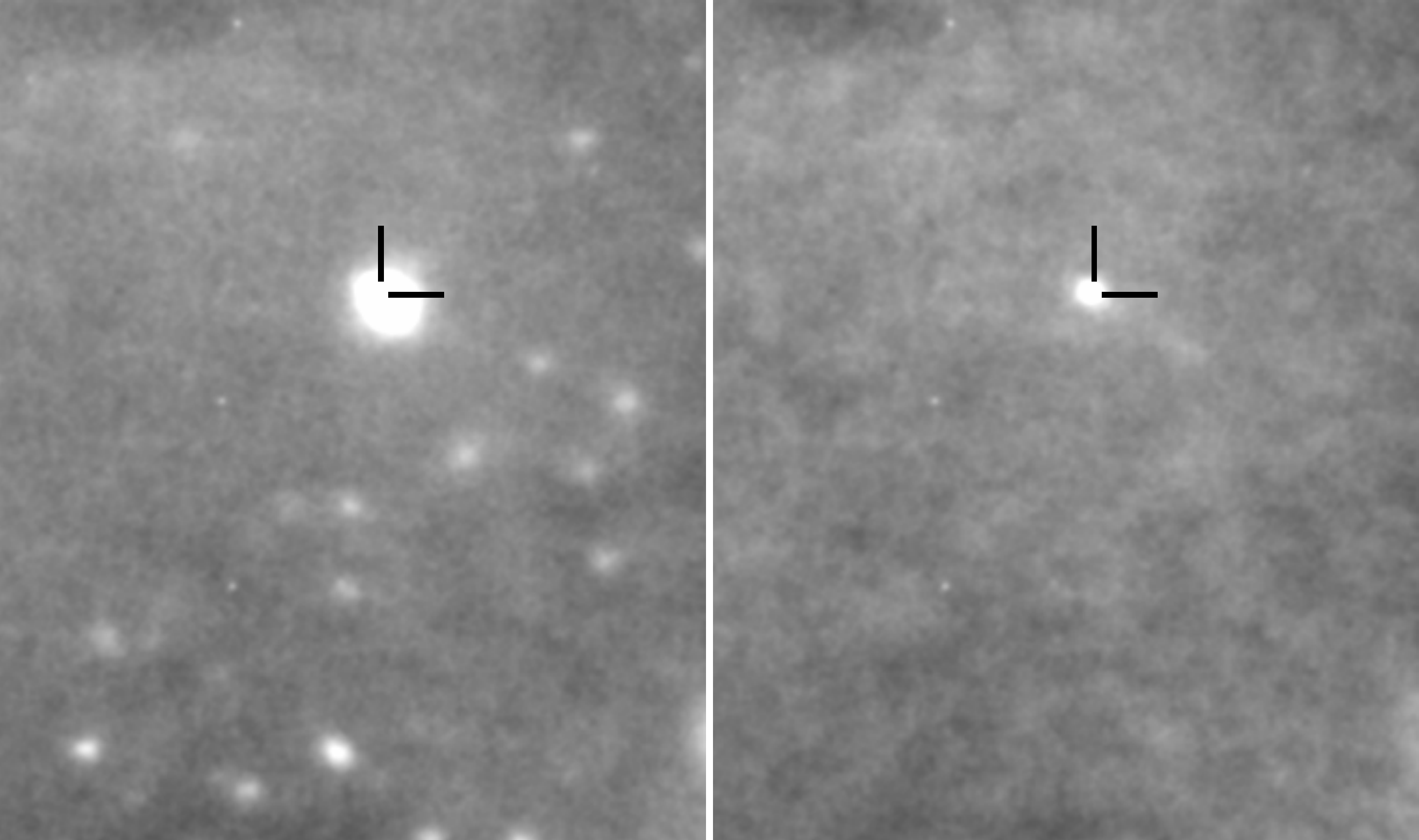}
 \caption{
Difference imaging of ZTF18aalrxas in \textit{r}-band, resulting in a detection around $r=20.6$\,mag. The science frame has been degraded by the PSF of the SDSS reference \citep{FrSo2016}. There are visible multiplicative gradients in the background due to the lack of flat fielding at the time the image was taken, which was the first night the instrument was on the telescope.}
 \label{fig:ZTF18aalrxas}
\end{figure}

With its robotic system and array of filters, KPED has the potential to contribute to the follow-up of objects found by all-sky surveys such as ZTF.
For objects not necessarily in the high cadence fields, or at depths not easily achieved by ZTF's nominal exposure times, dedicated follow-up is useful. 
As an example, we highlight ZTF18aalrxas, which is a SN IIb with a double-peaked lightcurve recently identified by ZTF (and will be described fully in a separate forthcoming publication).
Using difference imaging with ZTF images, a detection is made, reaching a depth around $r=22$\,mag.
This object highlights the ability to effectively perform deep difference imaging using survey images as templates, which will be useful for transient follow-up.

\section{Conclusion}
\label{sec:conclusion}

As we have discussed, there is significant scientific value for a camera with rapid readout like KPED on a robotic telescope. With the dynamical timescales of binary compact objects on the order of seconds, the ability to do high cadence photometry with systems such as KPED is essential. In addition, the measurements of eclipses, transits and occultations, which also occur on these time-scales, play an important role in constraining the physical parameters of the systems under study.

There are a number of potential routes to improve KPED. While the filter wheel allows for cycling between filters in order to get multi-wavelength information, these measurements are both observationally inefficient and are unable to resolve color variations on the time-scales shorter than constrained by the changing of filters. 
One possible solution is to use multiple channels, where multiple passbands are imaged simultaneously, which other systems have used to great effect \citep{HaHa2016,DhMa2007,DhMa2016}. 

In time-domain astronomy, spectroscopy is essential to classify transients discovered by wide FOV instruments. 
Due to the large sky localization areas of gravitational-waves, neutrino events, and some gamma-ray bursts, the number of optical candidates can vary from dozen to hundreds. As these sources are usually rapidly fading, reducing follow-up times for objects is one of the key requirements to face the challenge of time domain astronomy. A robotic telescope allows for rapid response.

Next, it is useful for constraining parameters of eclipsing white dwarf binaries \citep{HeUk2017}.
For systems with eclipses of order 30\,s, spectra composed of averages of that duration are required, otherwise other portions of the signal will be averaged. On the other hand, the EMCCD with the frame-transfer capability is capable of 1\,Hz (and faster) readouts, and therefore spectra can be phase-folded so as to get resolution on this scale.
In addition, because spectra naturally disperse the light over a wider range of pixels, resulting in lower signal-to-noise in any given pixel, the low light levels are ideal for the use of EMCCDs.

KPED reduction code is available at:\\
\url{https://github.com/dmitryduev/archiver-kped} and\\
\url{https://github.com/mcoughlin/kp84}.

\acknowledgments
MC is supported by the David and Ellen Lee Postdoctoral Fellowship at the California Institute of Technology.
The KPED team thanks the National Science Foundation, the National Optical Astronomical Observatory and the Murty family for support in the building and operation of KPED. In addition, they thank the CHIMERA project for use of the EMCCD.
This work has made use of data from the European Space Agency (ESA) mission {\it Gaia} (\url{https://www.cosmos.esa.int/gaia}), processed by the {\it Gaia} Data Processing and Analysis Consortium (DPAC, \url{https://www.cosmos.esa.int/web/gaia/dpac/consortium}). Funding for the DPAC has been provided by national institutions, in particular the institutions participating in the {\it Gaia} Multilateral Agreement.

\bibliographystyle{aasjournal}
\bibliography{references}

\end{document}